\documentclass[apj]{emulateapj}
\usepackage{amsmath}
\eqsecnum

\renewcommand\email\texttt

\newcommand\kms{{\;{\rm kms}^{-1}}}

\def\spose#1{\hbox to 0pt{#1\hss}}
\def\lta{\mathrel{\spose{\lower 3pt\hbox{$\sim$}}
    \raise 2.0pt\hbox{$<$}}}
\def\gta{\mathrel{\spose{\lower 3pt\hbox{$\sim$}}
    \raise 2.0pt\hbox{$>$}}}
\def\magsq{{\,\rm mag\, arcsec}^{-2}}

\begin{document} 

\slugcomment{\sc submitted to \it The Astrophysical Journal}
\shorttitle{\sc A Quintet of New Milky Way Companions} 
\shortauthors{}

\title{Cats and Dogs, Hair and A Hero: A Quintet of New Milky Way
Companions\footnotemark[$\dagger$]}

\author{V.\ Belokurov\altaffilmark{1},
D.\ B.\ Zucker\altaffilmark{1}, 
N.\ W.\ Evans\altaffilmark{1},
J.\ T.\ Kleyna\altaffilmark{2}, 
S.\ Koposov\altaffilmark{3},
S.\ T.\ Hodgkin\altaffilmark{1},
M.\ J.\ Irwin\altaffilmark{1},
G.\ Gilmore\altaffilmark{1},
M.\ I.\ Wilkinson\altaffilmark{1},
M.\ Fellhauer\altaffilmark{1},
D.\ M.\ Bramich\altaffilmark{1},
P.\ C.\ Hewett\altaffilmark{1},
S.\ Vidrih\altaffilmark{1}, 
J.\ T.\ A.\ De Jong\altaffilmark{3},
J.\ A.\ Smith\altaffilmark{4,5},
H.-W.\ Rix\altaffilmark{3},
E.\ F.\ Bell\altaffilmark{3},
R.\ F.\ G. Wyse\altaffilmark{6},
H.\ J.\ Newberg\altaffilmark{7},
P.\ A.\ Mayeur\altaffilmark{7,8},
B.\ Yanny\altaffilmark{9},
C.\ M.\ Rockosi\altaffilmark{10},
O.\ Y.\ Gnedin\altaffilmark{11},
D.\ P.\ Schneider\altaffilmark{12},
T.\ C.\ Beers\altaffilmark{13},
J.\ C.\ Barentine\altaffilmark{14},
H.\ Brewington\altaffilmark{14},
J.\ Brinkmann\altaffilmark{14},
M.\ Harvanek\altaffilmark{14},
S.\ J.\ Kleinman\altaffilmark{15},
J.\ Krzesinski\altaffilmark{14,16},
D.\ Long\altaffilmark{14},
A.\ Nitta\altaffilmark{17},
S.\ A.\ Snedden\altaffilmark{14}
}

\altaffiltext{1}{Institute of Astronomy, University of Cambridge,
Madingley Road, Cambridge CB3 0HA, UK;\email{vasily,zucker,nwe@ast.cam.ac.uk}}
\altaffiltext{2}{Institute for Astronomy, University of Hawaii, 2680
  Woodlawn Drive, Honolulu, HI 96822}
\altaffiltext{3}{Max Planck Institute for Astronomy, K\"{o}nigstuhl
17, 69117 Heidelberg, Germany}
\altaffiltext{4}{Los Alamos National Laboratory, ISR-4, MS D448, Los Alamos, NM 87545}
\altaffiltext{5}{Department of Physics and Astronomy, Austin Peay State University, P.O. Box 4608, Clarksville, TN 37040} 
\altaffiltext{6}{The Johns Hopkins University, 3701 San Martin Drive,
Baltimore, MD 21218}
\altaffiltext{7}{Rensselaer Polytechnic Institute, Troy, NY 12180}
\altaffiltext{8}{Department of Physics, Louisana Technical University,
Ruston, LA 71272}
\altaffiltext{9}{Fermi National Accelerator Laboratory, P.O. Box 500,
Batavia, IL 60510}
\altaffiltext{10}{Lick Observatory, University of California, Santa Cruz, CA 95064}
\altaffiltext{11}{Department of Astronomy, Ohio State University, 140 
West 18th Avenue, Columbus, OH 43210}
\altaffiltext{12}{Department of Astronomy and Astrophysics,
Pennsylvania State University, 525 Davey Laboratory, University Park,
PA 16802}
\altaffiltext{13}{Department of Physics and Astronomy, CSCE: Center for
the Study of Cosmic Evolution, and JINA: Joint Institute for Nuclear
Astrophysics, Michigan State University, East Lansing, MI 48824}
\altaffiltext{14}{Apache Point Observatory, P.O. Box 59, Sunspot, NM
88349}
\altaffiltext{15}{Subaru Telescope, 650 N. A'ohoku Place, Hilo, HI 96720}
\altaffiltext{16}{Mt.\ Suhora Observatory, Cracow Pedagogical University, ul.\ Podchorazych 2, 30-084 Cracow, Poland}
\altaffiltext{17}{Gemini Observatory, 670 N. A'ohoku Place, Hilo, HI 96720}

\footnotetext[$\dagger$]{Based in part on data collected at Subaru
Telescope, which is operated by the National Astronomical Observatory
of Japan.}

\begin{abstract}
We present five new satellites of the Milky Way discovered in Sloan
Digital Sky Survey (SDSS) imaging data, four of which were followed-up
with either the Subaru or the Isaac Newton Telescopes.  They include
four probable new dwarf galaxies -- one each in the constellations of
Coma Berenices, Canes Venatici, Leo and Hercules -- together with one
unusually extended globular cluster, Segue 1. We provide distances,
absolute magnitudes, half-light radii and color-magnitude diagrams for
all five satellites. The morphological features of the color-magnitude
diagrams are generally well described by the ridge line of the old,
metal-poor globular cluster M92. In the last two years, a total of ten
new Milky Way satellites with effective surface brightness $\mu_v
\gtrsim 28 \magsq$ have been discovered in SDSS data. They are less
luminous, more irregular and appear to be more metal-poor than the
previously-known nine Milky Way dwarf spheroidals. The relationship
between these objects and other populations is discussed. We note that
there is a paucity of objects with half-light radii between $\sim 40$
pc and $\sim 100$ pc. We conjecture that this may represent the
division between star clusters and dwarf galaxies.
\end{abstract}

\keywords{galaxies: dwarf -- Local Group}

\section{Introduction}

The known satellite galaxies of the Milky Way all lie within $\sim
300$ kpc and their brightest stars are resolvable from ground-based
telescopes. So, it is possible to acquire an enormous wealth of data
on their stellar populations, making the satellite galaxies important
objects in many fields of astrophysics~\citep[see
e.g.,][]{Do97,Sh03,To04,Pr05}. They have also emerged as a
battle-ground in Near-Field Cosmology~\citep{Fr02}. A fundamental
prediction of cold dark matter (CDM) theories is an abundance of
substructure in the non-linear regime. As noted by \citet{Kl99} and
\citet{Mo99}, galaxy assembly in CDM cosmogonies typically yields an
order of magnitude more dark haloes than there were known satellites
around the Milky Way.

Prior to the Sloan Digital Sky Survey~\citep[SDSS;][]{Yo00}, there
were 9 widely-accepted Milky Way dwarf spheroidals (dSphs), namely
Draco, Ursa Minor, Fornax, Carina, Sculptor, Leo I, Leo II, Sextans
and Sagittarius.  Seven of the Milky Way dSphs were discovered by eye
using photographic plates. The eighth, Sextans, was found by
\citet{Ir90} as part of a search of automated scans of photographic
plates, whilst the ninth, Sagittarius, was first identified
kinematically from radial velocity surveys of the Galactic
bulge~\citep{Ib95}.  The number of known Milky Way dSph satellites had
been increasing at a rate of one or two per decade before the advent
of SDSS.

The impact of SDSS has been dramatic. Four new Milky Way dSph
satellites have been discovered in SDSS data in quick succession: Ursa
Major~\citep{Wi05a}, Canes Venatici~\citep{Zu06a},
Bo{\"o}tes~\citep{Be06b} and Ursa Major II~\citep{Zu06c,Gr06},
together with what appears to be an unusually extended globular
cluster~\citep{Wi05b}. None are apparent in SDSS images, but all are
very clearly identifiable as overdensities of resolved stellar
objects.  This paper presents a further five new satellites found in
SDSS data, one each in the constellations of Coma Berenices, Canes
Venatici, Hercules, and two in Leo. We have confirmed four of these
discoveries with follow-up photometry on the Subaru telescope on Mauna
Kea and the Isaac Newton Telescope on La Palma. This brings the total
number of Milky Way companions found with SDSS data to ten, eight of
them probable dSphs.  This roughly doubles the number known prior to
SDSS.  They have eluded previous discovery because they are all of low
surface brightness ($\mu_v \gtrsim 28 \magsq$).

In fact, recent years have seen the discovery of a number of objects
that blur the hitherto clear distinction between star clusters and
dwarf galaxies. These include the ultra compact dwarf galaxies in the
Fornax cluster~\citep[e.g.,][]{Hi99,Dr00,Mi02}, the globular clusters
with unusually large half-light radii in M31~\citep{Hu05} and the
faint dSphs around M31~\citep{Zu06b,Ma06}. The 10 new SDSS discoveries
all lie in this poorly charted territory, where -- in the absence of
kinematic data -- the distinction between star clusters and dwarf
galaxies is hazy.

The paper is organised as follows: \S 2 provides a summary of the SDSS
and follow-up photometry on our 5 new discoveries, together with a
table of their properties.  \S3 reviews the relationship between
globular clusters and dwarf galaxies in the light of our new data, and
considers the implications of our discoveries for Near-Field
Cosmology. \S 4 summarizes our conclusions.

\begin{figure*}[t]
\begin{center}
\includegraphics[height=17cm]{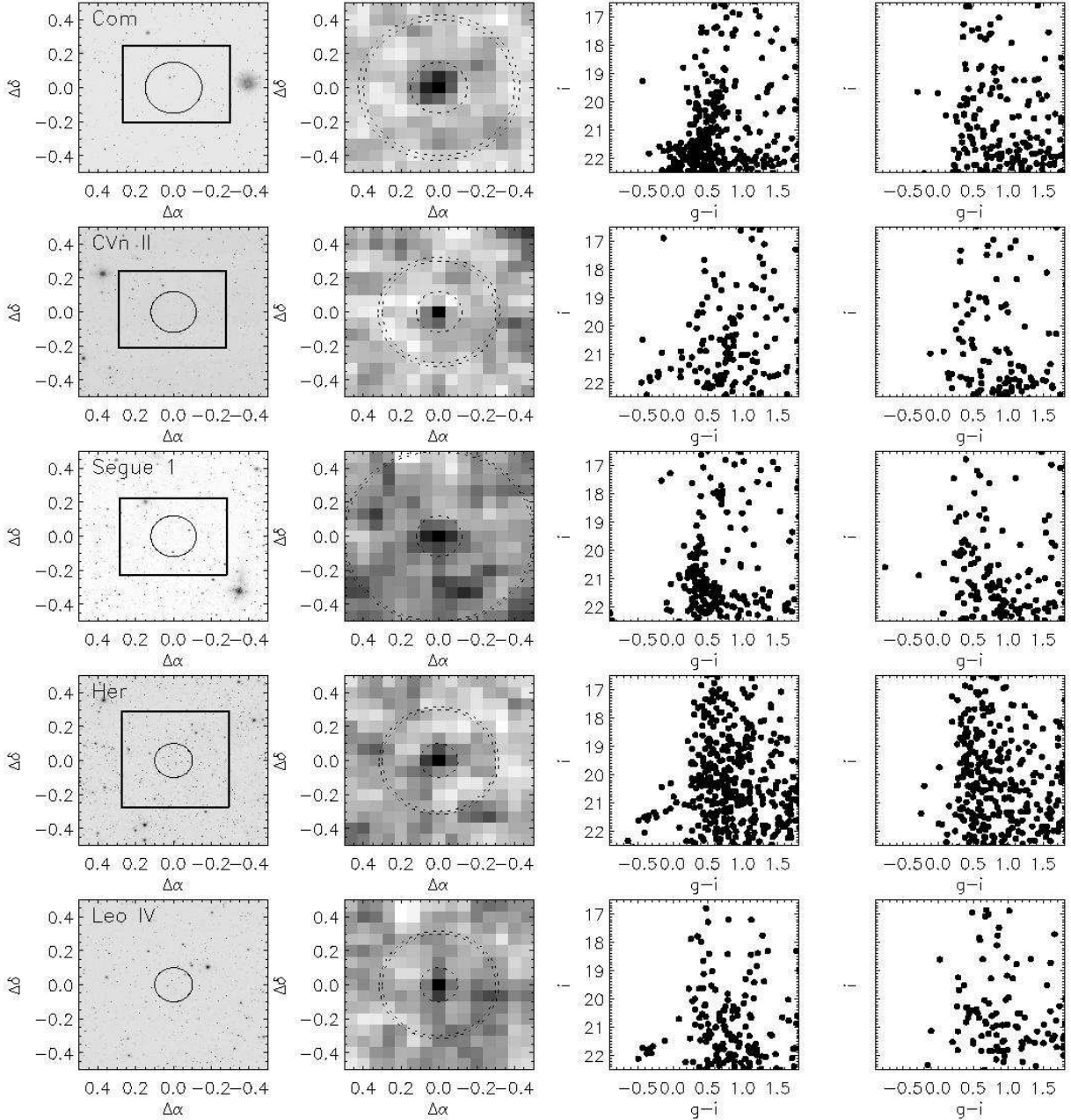}
\caption{Discovery panels for the 5 new satellites. The first column
is a cut-out of the SDSS, with a box showing the location of the
Subaru field ($34^\prime \times 27^\prime$) or INT field ($34^\prime \times 34^\prime$) and a circle marking the
central part of the object. The second column shows the pixellated
stellar density. The pixels are $4^\prime$ on each side. For each
object, 3 circles are shown of radii $r_1, r_2$ and $r_3$. The CMD of
stars lying within a circle of radius $r_1$ is given in the third
column. The CMD of stars lying in the annulus defined by the outer
radii ($r_2$ and $r_3$) is given in the fourth column. [$r_1,r_2,r_3$
for Coma are $0.15^\circ,0.4^\circ, 0.43^\circ$, for CVn II
($0.12^\circ, 0.3^\circ, 0.32^\circ$), for Segue 1 ($0.12^\circ,
0.5^\circ, 0.51^\circ$), for Her ($0.1^\circ, 0.3^\circ, 0.32^\circ$)
and for Leo IV ($0.1^\circ, 0.3^\circ, 0.32^\circ$].
\label{fig:objsa}}
\end{center}
\end{figure*}
\begin{figure*}[t]
\begin{center}
\includegraphics[height=4.5cm]{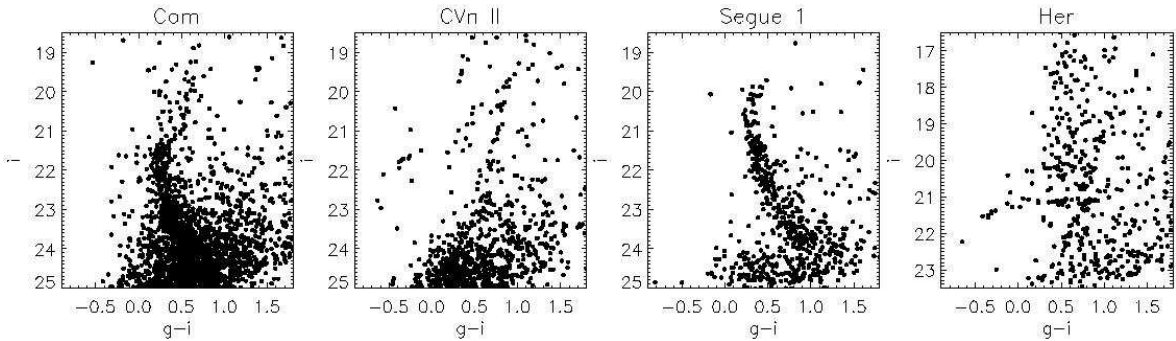}
\caption{CMDs of the central parts of Com, CVn II, Segue 1 and Her
from the Subaru/INT follow-up data.
\label{fig:extra}}
\end{center}
\end{figure*}
\begin{figure*}[t]
\begin{center}
\includegraphics[height=9cm]{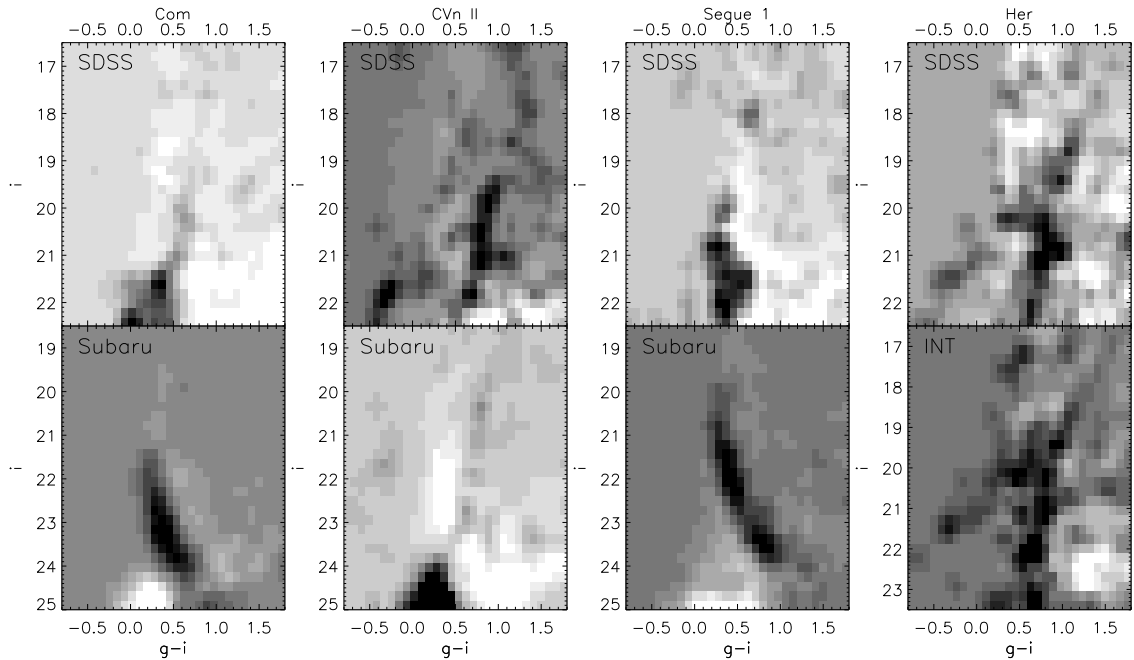}
\caption{Differential Hess diagrams using SDSS (upper panels) and
Subaru or INT (lower) data for Com, CVn II, Segue 1 and Her. In each
case, the normalized Hess diagram constructed with stars selected
within $r_1$ is subtracted from the normalized Hess diagram
constructed with stars selected between $r_2$ and $r_3$.
\label{fig:objsb}}
\end{center}
\end{figure*}
\begin{figure*}[t]
\begin{center}
\includegraphics[height=9cm]{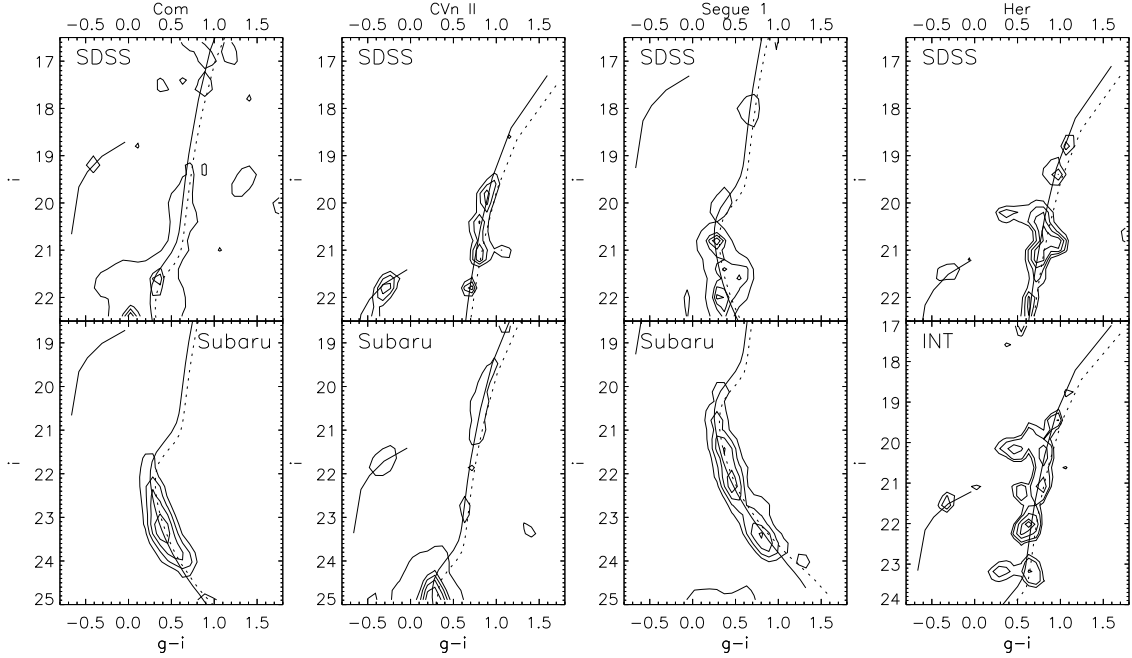}
\caption{Contour levels on the differential Hess diagrams using SDSS
(upper panels) and Subaru or INT (lower) data for Com, CVn II, Segue 1
and Her. The ridgeline of M92 (M13) is overlaid as a solid (dotted)
line, using the data of \citet{Cl05}.
\label{fig:objsc}}
\end{center}
\end{figure*}
\begin{figure*}[t]
\begin{center}
\includegraphics[height=7.5cm]{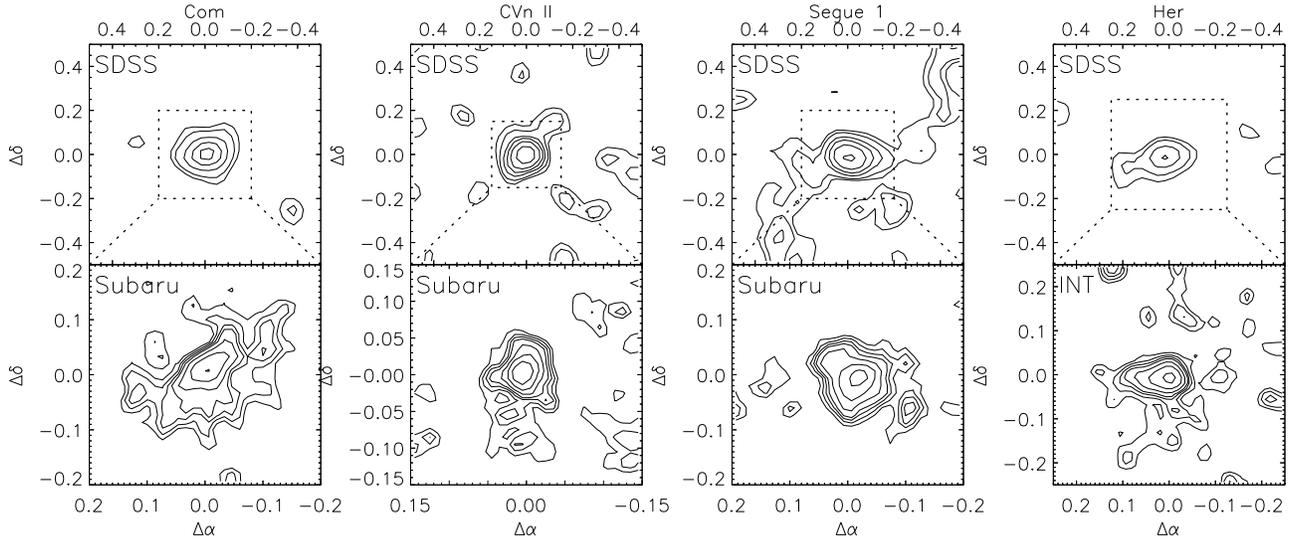}
\caption{Isodensity contours for Com, CVn II, Segue 1 and Her
I. Membership is determined using a mask constructed from the M92
ridgeline.  The top panels show CMD-selected stars with $18 < i <
22.5$. There are $30 \times 30$ pixels, smoothed
with a Gaussian with FWHM of 3 pixels.  Contour levels are 2,
3, 5, 7, 10, 15 $\sigma$ above the background. The bottom panels show the
central parts of the objects in Subaru/INT data.  There are $30 \times
30$ pixels, smoothed with a Gaussian with FWHM
of 2.2 pixels. Contour levels are 2, 3, 4, 5, 7, 10, 15 $\sigma$ above the
background.
\label{fig:objsd}}
\end{center}
\end{figure*}
\begin{figure*}[t]
\begin{center}
\includegraphics[height=5cm]{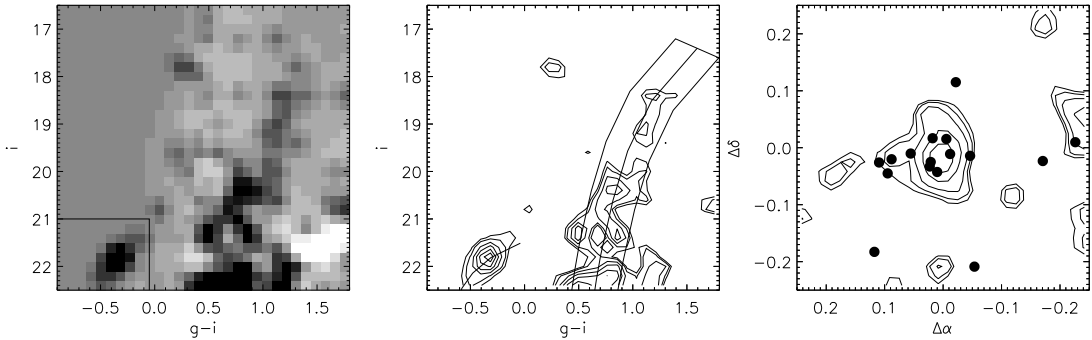}
\caption{Left: Differential Hess diagram for Leo IV, together with the
color-magnitude box used to select BHB candidate stars. Middle:
Contours of the differential Hess diagram, with overplotted M92
ridgeline and mask used to select members. Right: Isodensity contours
of Leo IV, together with locations of BHB candidate stars.
\label{fig:leo4}}
\end{center}
\end{figure*}

\begin{deluxetable*}{lccccc}
\tablecaption{Properties of the New Milky
Way Satellites \label{tbl:pars}} \tablewidth{0pt} \tablehead{
\colhead{Parameter\tablenotemark{a}} &  \colhead{Coma Berenices} &
\colhead{Canes Venatici II} & \colhead{Segue 1} &  \colhead{Hercules}
& \colhead{Leo IV} {~~~ } } \startdata  Coordinates (J2000) & 12:26:59
+23:54:15 &  12:57:10 +34:19:15 &  10:07:04 +16:04:55 &  16:31:02
+12:47:30 & 11:32:57 -00:32 00 \\ Galactic $(\ell,b)$ & $241.9^\circ$,
$83.6^\circ$  & $113.6^\circ$, $82.7^\circ$ & $220.5^\circ$,
$50.4^\circ$ & $28.7^\circ$, $36.9^\circ$ & $265.4^\circ$,
$56.5^\circ$ \\ Position Angle & $120^\circ$ & $0^\circ$ & $60^\circ$
& $125^\circ$  & $355^\circ$ \\ Ellipticity & $0.5$ & $0.3$ & $0.3$ &
$0.5$ & $0.25$ \\ $r_h$ (Plummer) & $5\farcm0$ & $3\farcm0$ &
$4\farcm5$ & $8\farcm0$ & $3\farcm3$\\   $r_h$ (Exponential) &
$5\farcm9$ & $3\farcm3$ & $4\farcm6$ & $8\farcm4$ & $3\farcm4$\\
V$_{\rm tot}$ & $14\fm5\pm0\fm5$ & $15\fm1\pm0\fm5$ & $13\fm8\pm0\fm5$
& $14\fm7\pm0\fm5$ & $15\fm9\pm0\fm5$ \\ (m$-$M)$_0$ & $18\fm2\pm0\fm2$ &
$20\fm9\pm0\fm2$ &$16\fm8\pm0\fm2$ &  $20\fm7\pm0\fm2$ &
$21\fm0\pm0\fm2$ \\ Heliocentric distance & $44^{+4}_{-4}$ kpc &
$150^{+15}_{-13}$ kpc  & $23^{+2}_{-2}$ kpc & $140^{+13}_{-12}$ kpc &
$160^{+15}_{-14}$ kpc\\ M$_{\rm tot,V}$ & $-3\fm7\pm0\fm6$ &
$-4\fm8\pm0\fm6$ & $-3\fm0\pm0\fm6$ & $-6\fm0\pm0\fm6$ &
$-5\fm1\pm0\fm6$ \enddata \tablenotetext{a}{Integrated magnitudes are
corrected for the Galactic foreground reddening reported by
\citet{Sc98}}
\label{tab:objs}
\end{deluxetable*}
\begin{figure*}
\begin{center}
\includegraphics[height=7cm]{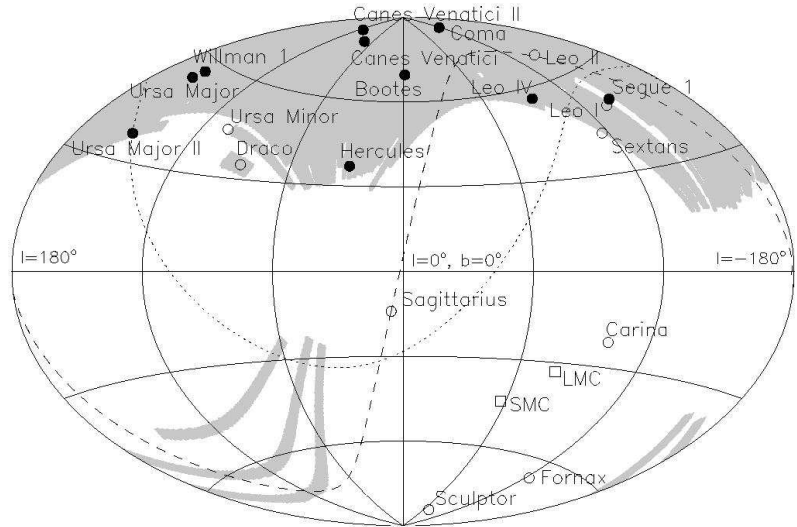}
\caption{The locations of Milky Way satellites in Galactic
coordinates.  Filled circles are satellites discovered by SDSS,
unfilled circles are previously known Milky Way dSphs. The light grey
shows the area of sky covered by the Sloan survey and its extensions
to date.  The dashed and dotted lines show the orbital planes of the
Sagittarius and Orphan Streams, respectively, taken from \citet{Fe06a}
and \citet{Fe06b}.}
\label{fig:dwarfslb}
\end{center}
\end{figure*}

\section{Discovery and Follow-Up}

\subsection{Data Acquistion and Analysis}

SDSS imaging data are produced in five photometric bands, namely $u$,
$g$, $r$, $i$, and $z$~\citep{Fu96,Gu98,Ho01,Am06,Gu06}.  The data are
automatically processed through pipelines to measure photometric and
astrometric properties \citep{Lu99,St02,Sm02,Pi03,Iv04,Tu06}. For
de-reddening, we use the maps of~\citet*{Sc98}. Data Release 5 (DR5)
primarily covers $\sim 8000$ square degrees around the North Galactic
Pole (NGP).  A small fraction of SDSS imaging data is not included in
DR5 and will be part of the future SDSS II/SEGUE~\citep{Ne03} data
release. All our satellites bar one (Segue 1) lie in DR5.

Here, we present further results from our ongoing systematic search
for Milky Way satellites using a variant of the algorithm described in
\cite{Be06b}. We experimented with a number of color cuts, pixel
binning and running window sizes in order to detect potential stellar
overdensities. The bins that were more than $4 \sigma$ away from the
background were selected and ranked according to statistical
significance. Visual inspection discarded a few obvious contaminants,
such as resolved stellar associations in background galaxies, on the
basis of their color-magnitude diagrams.

Figure~\ref{fig:objsa} shows 5 sets of 4 panels each derived from the
SDSS data. Each row refers to a different satellite. For ease of
exposition, it is helpful to have a simple name to call each object.
Even though the nature of these objects is at outset still to be
established, we will call those objects we believe to be dwarf
galaxies after their constellations, and those objects we believe to
be globular clusters after the survey. This nomenclature accords with
historical precedent.

The first row of Figure~\ref{fig:objsa} refers to a satellite in Coma
Berenices (henceforth Com), the second to a satellite in Canes
Venatici (henceforth CVn II), the third to a probable globular cluster
(henceforth Segue 1), the fourth to a satellite in Hercules
(henceforth Her) and the fifth to a satellite in Leo (henceforth Leo
IV)~\footnote{There is already a dwarf galaxy known in the
constellation of Canes Venatici~\citep{Zu06a}, and two known in
Leo. Leo III is an alternative name for the Leo dwarf irregular, also
called Leo A~\citep{Be00}.}. For each object, the first column
provides a grey-scale image centered on the satellite; no obvious
objects can be seen. The second column is a density map of all the
objects classified by the SDSS pipeline as stars; a stellar
overdensity is visible in the center of each plot.  In each case, an
inner circle and an outer annulus are shown in dotted lines. The third
and fourth columns show color-magnitude diagrams (CMDs) constructed
from all stars in the central region and in the annulus,
respectively. We will shortly use these to construct Hess diagrams,
but first we describe additional data acquired on 4 of the 5
satellites.

Deeper follow-up observations of Com, CVn II and Segue 1 were made at
the Subaru telescope on Mauna Kea, using the Suprime-Cam mosaic
camera.  The data were gathered on 2006 May 25 (UT), using a single
pointing to cover each stellar overdensity. In each case, the location
of the Subaru field is shown in the panels in the first column of
Figure~\ref{fig:objsa}. Each pointing was observed in $g'$ and $i'$
bands in a 3-exposure dither to cover the gaps between CCDs. For ease
of comparison, the Subaru $g',i'$ photometry was boot-strapped onto
the SDSS $g,i$ photometric system. The Subaru data, although
restricted to the central parts, are roughly 2.5 magnitudes deeper in
the $i$ band.  Further details on the Subaru data acquisition and
processing are given in \citet{Zu06c}.

Follow-up photometric observations of Her were made with the 2.4m
Isaac Newton Telescope on the island of La Palma on the night of 2006
June 27 (UT). Images were taken with the prime focus Wide-Field
Camera, which has a footprint of $34 \times 34$ arcminutes and a pixel
scale of 0.33 arcseconds. Exposures comprised three dithered 600
second integrations in each of the $g'$ and $i'$ filters (for a total
of 30 minutes of exposure in each filter). Data were processed using a
general purpose pipeline for processing wide-field optical CCD data
described elsewhere~\citep{Ir01} and boot-strapped onto the SDSS
photometric system. The INT data are roughly a magnitude deeper in the
$i$ band than the SDSS data.

Figure~\ref{fig:extra} shows the CMDs of the central parts (marked on
Figure~\ref{fig:objsa} as circles) of Com, CVn II, Segue 1 and Her
using the follow-up data.  The upper panels of Figure~\ref{fig:objsb}
show the difference between the normalized color-magnitude diagrams of
the inner and outer parts of Com, CVn II, Segue 1 and Her in SDSS data
(the third and fourth columns of Figure~\ref{fig:objsa}). The lower
panels show the same physical quantity, but this time constructed with
the deeper data from Subaru/INT. In these differential Hess diagrams,
familiar features, such as giant branches, horizontal branches and
upper main sequences, can all be discerned. We conclude that each of
the 4 objects is localized and has a distinct stellar population. This
provides reassuring confirmation that these 4 objects are new
satellites.

For the purpose of clarity, the Hess diagrams are converted to contour
plots in Figure~\ref{fig:objsc}. The ridgelines of the Galactic
globular clusters M92 ([Fe/H] $\sim -2.24$) and M13 ([Fe/H] $\sim
-1.65$) are overlaid, using the data of \citet{Cl05} transformed into
the SDSS photometric system. The ridgeline of very old, metal poor
M92 gives a remarkably good representation of the stellar populations.
In the cases of Com and Segue 1, the main sequence and giant branch
can be matched well. In the cases of CVn II and Her, the turn-off,
giant branch and, most importantly, the horizontal branch are
well-fit. This comparison immediately gives us a good estimate of the
distance modulus to each object, as listed in
Table~\ref{tab:objs}. Segue 1 and Com are reasonably close, at
heliocentric distances of $\sim 23$ kpc and $\sim 44$ kpc, whilst CVn
II and Her are further away at distances of $\sim 150$ kpc and $\sim 140$ kpc,
respectively. To define membership of each object, we use the M92
ridgeline to construct a mask at the estimated
distance. Figure~\ref{fig:objsd} shows the isodensity contours of
stars matching the mask for each object using SDSS and follow-up
data. They are all extended and rather irregular in their outer
parts. Com is the closest and has the most substructure. CVn II and
Her are rounder, but there is evidence for extensions that may be part
of streams or tails. Segue 1 is the smallest. Its innermost contours
are quite round, but there is clearly a tail visible in the SDSS data.

We have no follow-up data for Leo IV. However, its CMD, shown in
Figure~\ref{fig:leo4}, reveals a giant branch and blue horizontal
branch. As before, the ridgeline of M92 gives a reasonable match (see
middle panel of Figure~\ref{fig:leo4}), but the width of the giant
branch appears to be larger than that of a single stellar population. The
isodensity contours are shown in the right panel. Black dots indicate
candidate blue horizontal branch stars. It is reassuring to see that they are
concentrated and extended in the same manner as the isodensity
contours. In the absence of follow-up data, we regard this as a useful
check.

A number of integrated photometric and morphological parameters for
Com, CVn II, Segue 1, Her and Leo IV are reported in
Table~\ref{tab:objs}. The algorithms for the calculations of position
angle, ellipticity, half-light radius, and absolute magnitude are
described in detail in our earlier papers~\citep{Zu06a,Be06b}.

\subsection{Summary of the New Satellites}

Based on their sizes and shapes, our working hypothesis is that Com, CVn
II, Her, and Leo IV are new dwarf galaxies, whilst Segue 1 is an
extended globular cluster.

{\it Com} is located at a heliocentric distance of $44 \pm 4$ kpc. It
has a half-light radius of $\sim 70$ pc, although this may be an
underestimate given its irregular and extended shape. Its CMD is
consistent with that of a single, old stellar population of
metallicity [Fe/H] $\sim -2$.

{\it CVn II} is at a distance of $150^{+15}_{-14}$ kpc and has a
half-light radius of $\sim 140$ pc. The central density contours are
round, but there is a southward extension clearly visible in the deep
Subaru data. Its CMD has a clearly defined subgiant branch with a hint
of a red clump, a reasonably prominent blue horizontal branch and a
narrow giant branch.

{\it Segue 1} is the closest at a distance of $23 \pm 2$ kpc. Its
half-light radius is 30 pc, roughly the same size as the largest Milky
Way globular clusters, such as Pal 5 and Pal 14~\citep{Ha96}. There is
evidence for tidal tails in the SDSS data. Its CMD has a poorly
populated subgiant branch and no obvious horizontal branch. At $\alpha
\approx 152^\circ, \delta \approx 16^\circ$, Segue 1 is superposed on
the Sagittarius stream; at this location \citet{Be06a} estimated the
distance to the Sagittarius stream to be $\sim 20$ kpc, close to the
distance to Segue 1. We conclude that Segue 1 is likely a globular
cluster formerly associated with the Sagittarius dSph.

{\it Her} lies at a distance of $140^{+13}_{-12}$ kpc and has a half-light
radius of $\sim 320$ pc. It has an extended morphology. Its CMD shows
not just a giant branch, but both blue and red horizontal branches,
which may hint at multiple stellar populations.

{\it Leo IV} is at a distance of $160^{+15}_{-14}$ kpc. Its half-light
radius is $\sim 160$ pc. Its CMD is more complex than the others, with
an apparent thick giant branch and a blue horizontal branch. The thickness may
be caused by multiple stellar populations and/or by depth along the
line of sight.

\begin{figure*}
\begin{center}
\includegraphics[height=11cm]{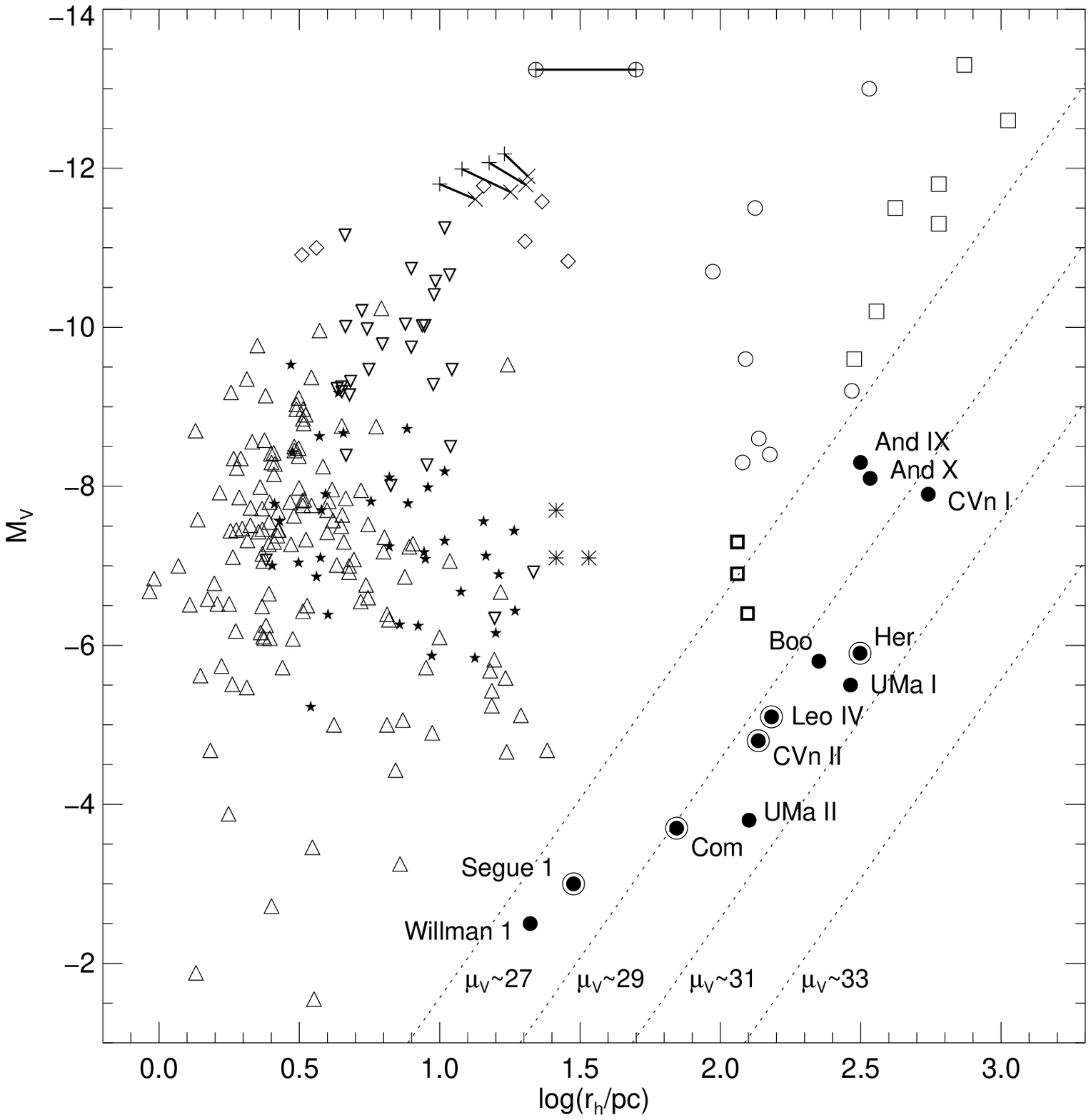}
\caption{Location of different classes of object in the plane of
absolute magnitude versus half-light radius.  Lines of constant
surface brightness are marked.  Filled circles are the SDSS
discoveries including the 10 Milky Way
satellites~\citep{Wi05a,Wi06,Zu06a,Zu06c,Be06b}, as well as And IX and
X~\citep{Zu04,Zu06b}.  Unfilled circles are eight previously known
Milky Way dSphs with Sgr omitted~\citep{Ir95,Ma98}, squares are the
M31 dSphs \citep{Mc06}, bold squares are three new M31 dSphs recently
discovered by \citet{Ma06}, while triangles are the Galactic globular
clusters \citep{Ha96}. A variety of other extragalactic objects are also
plotted: asterisks are the extended M31 globular clusters discovered
by \citet{Hu05}, pluses and crosses are UCDs in Fornax from
\citet{Mi02} and \citet{DeP05} respectively, diamonds are the
so-called Virgo dwarf-globular transition objects~\citep{Ha05}, while
filled stars and inverted triangles are globular clusters from the
nearby giant elliptical NGC 5128 from \citet{Ha02} and \citet{Go06}
respectively. Different measurements of the same object are connected
by straight lines. The straight line connecting the Earth symbols
refer to measurements by \citet{Mi02} and \citet{Dr03} of UCD3 in
Fornax.}
\label{fig:magversusrh}
\end{center}
\end{figure*}

\section{Discussion}

\subsection{Dwarf Galaxies or Globular Clusters}

The five objects in this paper, together with the five Milky Way
satellites previously discovered in SDSS data -- namely Ursa Major I,
Willman 1, Canes Venatici I, Bootes and Ursa Major II -- can be
usefully taken together as a group. They were all discovered in the
same dataset with similar methods, although this does not necessarily
imply any underlying physical commonality. The locations of the ten
SDSS objects in the Galactic sky are shown in
Figure~\ref{fig:dwarfslb}, together with the 9 previously known
dSphs. Prior to SDSS, it had long been suspected that there may be
some missing dSphs at low Galactic latitude in the Zone of
Avoidance~\citep[see e.g.][]{Ma98}. However, the SDSS objects all lie
at high Galactic latitude, as the survey is concentrated around the
North Galactic Pole. It is difficult to escape the conclusion that
there are many more Milky Way companions waiting to be
discovered. Assuming (i) all dwarf satellites in the area of sky
covered by SDSS have been found and (ii) the distribution of dwarf
satellites is isotropic, then there may be $\sim 50$ dwarfs in all. In
fact, both assumptions are surely incorrect. Systematic surveys for
all satellites in SDSS DR5 are underway (e.g., Koposov et al. 2006, in
preparation) and will undoubtedly uncover further candidates. The
spatial distribution of dwarf galaxies is a controversial issue,
although the most recent analysis of the simulation data suggests that
dwarf satellites may lie preferentially along the major axis of the
light distribution of the host galaxy~\citep[see e.g.,][and references
therein]{Ze05,Ya06}. If so, then our extrapolation to a total of $\sim
50$ dwarfs may still be a underestimate.

Figure~\ref{fig:magversusrh} shows objects plotted in the plane of
absolute magnitude and half-light radius. This includes the ten SDSS
discoveries in the Milky Way (filled circles) and the eight Milky Way
dSphs omitting Sgr (unfilled circles). We have added to the sample of
SDSS discoveries two dSphs found around M31, namely And IX and
X~\citep{Zu04,Zu06b}. Also shown are a number of populations of
extragalactic objects -- such as the M31 dSphs, including the most
recent 3 discoveries by~\citet{Ma06}, the three unusually extended
globular clusters found in M31 by~\citet{Hu05}, the ultra-compact
dwarf galaxies (UCDs) in the Fornax and Virgo
clusters~\citep{Mi02,DeP05,Ha05} and globular clusters from the nearby
giant elliptical NGC 5128~\citep{Go06,Ha02}. Some lines of constant
surface brightness are also marked. This shows why the recent spate of
discoveries in SDSS data is occurring -- the survey is reaching much
lower surface brightnesses than possible before. All the SDSS
discoveries lie below, and all the previously known Milky Way dSphs
above, the line marking $\mu_V =27 \magsq$.

Some properties of the SDSS discoveries are apparent from
Figure~\ref{fig:magversusrh}. As a group, they are much fainter than
the previously known Milky Way and M31 dSphs. They are also less
regular in shape, which suggests tidal effects may be important. Of
course, caution is needed, as some of the isophotal distortion is due
to low object counts and uncertain background
subtraction. Nonetheless, there seems to be a rough correlation
between irregularity and distance, as Boo, UMa II and Com are the most
irregular and also amongst the closest. They all seem to be very
metal-poor with [Fe/H] $\approx -2$, at least as judged by the fit of
M92's ridgeline to giant branch, main sequence turn-off or horizontal
branch of the CMDs. This is supported by the recent measurement of the
metallicity of Boo~\citep{Mu06} as [Fe/H] $\sim -2.5$. The SDSS
discoveries are larger and somewhat less luminous than typical Milky
Way globular clusters.

The seeming lack of metals in the SDSS discoveries is interesting. The
Galactic halo contains a significant fraction of stars more metal-poor
than [Fe/H] $\sim -2.0$~\citep[see e.g.,][]{Ch04,Be05}. The previously
known dSphs on the other hand contain very few metal-poor stars~\citep[see e.g.,][]{To04, Ko06}. The
new SDSS discoveries may be representatives of the population that
built the old, metal-poor component of the Milky Way halo.

Also apparent in Figure~\ref{fig:magversusrh} is the fact that the
data-points fall into a number of clumps. The Milky Way globulars form
one obvious grouping. A number of unusual objects, such as the
extended M31 clusters and the UCDs in Fornax and Virgo, all lie in
regions abutting the globulars in the plane of absolute magnitude and
half-light radius. For example, UCDs are brighter than Galactic
globulars, but they could be the bright tail of the globular cluster
systems in the Fornax and Virgo clusters. Separating the globulars
from the dwarf galaxies is a sparsely populated vertical band
corresponding to half-light radii between $\sim 40$ pc to $\sim 100$
pc. Only two objects lie in this gap. The first is Com, which is so
irregular that its half-light radius is susceptible to significant
uncertainties. The second is UCD3 in Fornax as measured by
\citet{Mi02}. A {\it Hubble Space Telescope} re-measurement of the
half-light radius of this object by \citet{Dr03} yielded a somewhat
smaller answer. The two measurements are connected by a straight line
in Figure~\ref{fig:magversusrh}.

The gap is suggestive, but not conclusive, as SDSS DR5 covers only 20
per cent of the night sky around the North Galactic Cap. There are
still very few objects in Figure~\ref{fig:magversusrh} at low surface
brightness.  However, it is significant that there are SDSS
discoveries on either side of the gap.  It is also clear that, if
there were a population of extended, luminous star clusters in the
Milky Way analogous to those found by \cite{Hu05} in M31, then they
would have very likely been found already in SDSS data. In this
picture, Segue 1 and Willman 1 are unusually faint, extended, globular
clusters, whilst the remaining SDSS discoveries are dwarf galaxies.
Of course, the separation between clusters and dwarf galaxies would be
much clearer on plots of absolute magnitude versus velocity
dispersion. It will be interesting to see analogues of
Figure~\ref{fig:magversusrh} once kinematic data become available.

At the moment, all objects to the left of the gap show no evidence of
dynamically significant dark matter. All the objects to the right with
measured kinematics are consistent with substantial amounts of dark
matter. For the classical dSphs, the kinematic data are consistent
with a common halo mass scale~\citep[e.g.,][]{Wil06}. This is also the case for
UMa~I \citep{Kl05}.  If this holds for all the new SDSS discoveries,
it might provide a natural explanation for the existence of a gap.


\subsection{Implications for Near-Field Cosmology}

The objects discussed in this paper have a number of implications for
Near-Field Cosmology.  In CDM, dark matter overdensities collapse to
form cusped halos, with the smallest and least massive halos being the
densest. The simulations of \citet{Kl99} and \citet{Mo99} predicted
hundreds of small Galactic satellite halos, as compared to the handful
of then known satellite galaxies around the Milky Way. If each small
dark matter halo indeed harbors a detectable small galaxy, then there
is a dramatic conflict between predictions and observations.  It
remains unclear whether theory or observations are responsible for
this discrepancy. In fact, many theoreticians responded to this result by
developing models that suppress gas accretion~\citep[see e.g.][]{Ef92}
or star formation in low mass halos. This produces a large population
of entirely dark satellites~\citep[see e.g.][]{Bu00,Kr04,Mo06},
together with a much smaller number of dSphs, roughly in accord with
the datum of 9 dSphs per large galaxy. However, it is now clear, from
the discoveries over the past couple of years, that the observational
situation has changed dramatically.

Spectroscopic studies are urgently needed to assess the dark matter
content of the SDSS discoveries. So far, only two of the galaxies have
kinematic data. \citet{Kl05} measured the velocities of seven UMa I
stars and obtained a velocity dispersion of $\sim 9 \kms$ and a
mass-to-light ratio of $\sim 500$.  \citet{Mu06} measured the radial
velocities of seven Boo stars and obtained a velocity dispersion of
$\sim 7 \kms$ and a mass-to-light ratio of between 130 and 680. Caution
is needed in interpreting these results as they are calculated under
the strong assumption of steady-state, virial equilibrium.  Based on
these results, UMa I and Boo would be the two most dark matter
dominated objects known in the Universe. The implication is that the
SDSS discoveries may well be members of the missing population of low
stellar mass, dark matter dominated galaxies originally predicted by
CDM. Only when a complete census of these objects has been obtained
will we be able to assess whether the properties of the population are
consistent with the predictions of the simulations.

Another possibility is that the Milky Way satellites condensed out of
the tidal tails of an early merger with a gas-rich
progenitor~\citep{Ba92}; this would make them analogous to the tidal
dwarf galaxies observed in interacting systems today~\citep{We00}. An
attractive facet of this idea is that it naturally accounts for
possible streams in the Milky Way dSphs. This phenomenon was
originally spotted by~\citet{Ly82a,Ly82b}, who noted that the bright
dSphs may be aligned in one or two streams of tidal debris. However,
examining Figure~\ref{fig:dwarfslb}, it is apparent that the simple
model of ~\citet{Kr05}, in which most of the Milky Way satellites are
associated with a single disk-like structure, is hard to reconcile
with the new data.

\citet{Kr97} has studied the long-time evolution of tidal dwarf
galaxies.  The idea is that tidal dwarf galaxies with no dark matter
suffer destruction at perigalacticon passages to leave orbiting but
unbound agglomerations of stars that appear compact near their
apocenter, and which constitute some of the present-day dSphs. The
absence of velocity gradients and the thinness of the horizontal
branch in galaxies like Draco~\citep{Kl02,Kl03}, Fornax and
Sagittarius~\citep{Ma03} means that this theory cannot reproduce the
observed properties of the brightest dSphs. However, the irregular
shape and the abundance of substructure in the objects presented by
\citet{Kr97} do bear a striking resemblance to the new SDSS
discoveries, although Kroupa's objects as a class are much more
luminous and may require fortuitous timing and a favorable viewing
angle. It would be interesting to see detailed predictions of the
properties of these objects at fainter absolute magnitudes ($M_V
\approx -6$).

\section{Conclusions}

In this paper, we announced the discovery of five new satellites of
the Milky Way. One is a probable new globular cluster, which has been
named Segue 1 after the survey in which it was found. It is in the
Sagittarius stream and was possibly stripped off the Sagittarius
progenitor.  The remaining four are probably new dwarf galaxies, which
have been named according to their constellations as Coma Berenices,
Canes Venatici II, Leo IV and Hercules. We have presented SDSS and
deeper Subaru/INT photometry (where available) on these objects. We
provide color-magnitude diagrams, distances, absolute magnitudes and
half-light radii for all the satellites.

Taken together with the earlier announcements of Ursa Major I, Willman
1, Canes Venatici I, Bootes and Ursa Major II, ten new Milky Way
satellites have been discovered in SDSS data in very rapid
succession. This abundance of discoveries is occurring because the
survey is probing down to hitherto uncharted surface brightnesses.
All the SDSS discoveries are effective surface brightness fainter than
$\mu_V =27 \magsq$. The obvious conclusion is that there are more low
surface brightness Milky Way satellites waiting to be discovered.

The SDSS discoveries occupy a distinct region in the plane of absolute
magnitude versus half-light radius. They are typically fainter, more
metal-poor and more irregular than the previously-known Milky Way
dwarf spheroidals (dSphs).  They are larger, and somewhat less
luminous, than typical Galactic globular clusters.  Taking the known
globular clusters, the previously known Milky Way satellites and the
SDSS discoveries, there is still a scarcity of objects with half-light
radii between $\sim 40$ pc and $\sim 100$ pc. This may represent the
division between star clusters and dwarf galaxies.  

The SDSS discoveries could have a bearing on the ``missing satellite''
problem.  Preliminary indications from studies of UMa I~\citep{Kl05}
and Boo~\citep{Mu06} suggest that these objects may be dark matter
dominated.  It seems possible that a population of ultra-faint, dark
dwarf galaxies really does surround the Milky Way. However, it is not
yet clear that these are the ``missing satellites'' predicted by
the simulations of \citet{Kl99} and \citet{Mo99}. The match of the
data to CDM halos should be carried out in the plane of cumulative
number versus halo mass or circular velocity. 

Kroupa's (1997) study of the evolution of tidal dwarf galaxies
reproduces some of the properties of the new satellites. This opens up
the possibility that some of these objects may be tidal dwarf
galaxies, or shreds from the violent building phase of the Milky Way.
In this case, the satellites will not have substantial dark
matter. Kinematic data are now urgently needed to confirm whether or
not these objects are dark matter dominated.

\acknowledgments 
We thank Pavel Kroupa and Beth Willman for useful discussions and
comments.  V.B., D.B.Z., M.I.W., M.F. and  D.M.B. acknowledge the
financial support of the Particle Physics and Astronomy Research
Council of the United Kingdom.

Funding for the SDSS and SDSS-II has been provided by the
Alfred P.  Sloan Foundation, the Participating Institutions, the
National Science Foundation, the U.S. Department of Energy, the
National Aeronautics and Space Administration, the Japanese
Monbukagakusho, the Max Planck Society, and the Higher Education
Funding Council for England. The SDSS Web Site is
http://www.sdss.org/.

The SDSS is managed by the Astrophysical Research Consortium for the
Participating Institutions. The Participating Institutions are the
American Museum of Natural History, Astrophysical Institute Potsdam,
University of Basel, Cambridge University, Case Western Reserve
University, University of Chicago, Drexel University, Fermilab, the
Institute for Advanced Study, the Japan Participation Group, Johns
Hopkins University, the Joint Institute for Nuclear Astrophysics, the
Kavli Institute for Particle Astrophysics and Cosmology, the Korean
Scientist Group, the Chinese Academy of Sciences (LAMOST), Los Alamos
National Laboratory, the Max-Planck-Institute for Astronomy (MPIA),
the Max-Planck-Institute for Astrophysics (MPA), New Mexico State
University, Ohio State University, University of Pittsburgh,
University of Portsmouth, Princeton University, the United States
Naval Observatory, and the University of Washington. This work was
based in part on observations made with the Isaac Newton Telescope on
the Island of La Palma by the Isaac Newton Group in the Spanish
Observatorio del Roque de los Muchachos of the Institutode Astrofisica
de Canarias.

\end{document}